\documentclass[acmtog, screen]{acmart}
\setcitestyle{square}

\usepackage{booktabs} % For formal tables
\usepackage{graphicx}
\usepackage{tabularx}
\usepackage{lineno}
\usepackage{float}
\usepackage{braket}
\usepackage{amsmath}
\usepackage{amssymb}
\usepackage{amsthm}
\usepackage{listings}
\usepackage{paralist}
\usepackage{booktabs}
\usepackage{lineno}
\usepackage{booktabs}
\usepackage{dsfont}
\usepackage{subcaption}
\usepackage{wrapfig}
\usepackage{booktabs,arydshln}
\usepackage{bm}
\usepackage{algorithm}% http://ctan.org/pkg/algorithms
\usepackage[noend]{algpseudocode}
\usepackage{mathtools}
\usepackage{multirow}
\usepackage{pifont}
\let\oldding\ding
\renewcommand{\ding}[2][1]{\scalebox{#1}{\oldding{#2}}}

\newcommand{\argmin}{\operatornamewithlimits{argmin}}

\newcommand{\myvec}[1]{{\mathbf #1}}

\newcommand\numberthis{\addtocounter{equation}{1}\tag{\theequation}}

\usepackage{hyperref}
\hypersetup{pdfstartview={XYZ null null 1.00}}
\hypersetup{colorlinks=true}
\hypersetup{citecolor=blue}
\hypersetup{linkcolor=blue}
\hypersetup{allcolors=blue}

\usepackage{color}
\usepackage{tikz} % after \usepackage{color}!
\usepackage{pgfplots}
\usepackage{pgfplotstable}
\pgfplotsset{compat=1.8}

\usepackage{xcolor}

\definecolor{red}{rgb}{0.8,0,0}
\definecolor{green}{rgb}{0.0,0.5,0}
\definecolor{gray}{rgb}{0.75,0.75,0.75}
\definecolor{blue}{rgb}{0.0,0.0,0.6}
\definecolor{mygrey}{rgb}{0.3,0.3,0.3}

\newcommand{\Revision}[1]{{#1}}

\makeatletter
\def\adl@drawiv#1#2#3{%
        \hskip.5\tabcolsep
        \xleaders#3{#2.5\@tempdimb #1{1}#2.5\@tempdimb}%
                #2\z@ plus1fil minus1fil\relax
        \hskip.5\tabcolsep}
\newcommand{\cdashlinelr}[1]{%
  \noalign{\vskip\aboverulesep
           \global\let\@dashdrawstore\adl@draw
           \global\let\adl@draw\adl@drawiv}
  \cdashline{#1}
  \noalign{\global\let\adl@draw\@dashdrawstore
           \vskip\belowrulesep}}
\makeatother

\setcopyright{rightsretained}
\setcopyright{rightsretained}
\begin{document}

\title{Photorealistic Material Editing Through Direct Image Manipulation}
%\titlenote{Produces the permission block, and
%  copyright information}
%\subtitle{Extended Abstract}
%\subtitlenote{The full version of the author's guide is available as
%  \texttt{acmart.pdf} document}

\author{K\'{a}roly Zsolnai-Feh\'{e}r}
%\author{Anonymous Author 1}
%\authornote{E-mail address: zsolnai@cg.tuwien.ac.at}
%\authornote{user@emailaddress.com}
\affiliation{%
  \institution{TU Wien}
%  \institution{Anonymous Institution}
  \streetaddress{Favoritenstrasse 9-11/193-02}
  \city{Vienna, Austria}
  \postcode{1040}
}
\email{zsolnai@cg.tuwien.ac.at}

\author{Peter Wonka}
%\author{Anonymous Author 2}
%\authornote{pwonka@gmail.com}
%\authornote{The secretary disavows any knowledge of this author's actions.}
\affiliation{%
  \institution{KAUST}
  %\institution{Anonymous Institution}
  \streetaddress{Al Khwarizmi Bldg 1}
  \city{Thuwal, Kingdom of Saudi Arabia} 
  \postcode{23955-6900}
}
\email{pwonka@gmail.com}

\author{Michael Wimmer}
%\author{Anonymous Author 3}
%\authornote{This author is the}
%\authornote{wimmer@cg.tuwien.ac.at}
\affiliation{%
  \institution{TU Wien}
%  \institution{Anonymous Institution}
  \streetaddress{Favoritenstrasse 9-11/193-02}
  \city{Vienna, Austria}
  \postcode{1040}
}
\email{wimmer@cg.tuwien.ac.at}
\renewcommand{\shortauthors}{Zsolnai-Feh\'{e}r et al. 2018}

\begin{abstract}
Creating photorealistic materials for light transport algorithms requires carefully fine-tuning a set of material properties to achieve a desired artistic effect. This is typically a lengthy process that involves a trained artist with specialized knowledge. In this work, we present a technique that aims to empower novice and intermediate-level users to synthesize high-quality photorealistic materials by only requiring basic image processing knowledge. In the proposed workflow, the user starts with an input image and applies a few intuitive transforms (e.g., colorization, image inpainting) within a 2D image editor of their choice, and in the next step, our technique produces a photorealistic result that approximates this target image. Our method combines the advantages of a neural network-augmented optimizer and an encoder neural network to produce high-quality output results within 30 seconds. We also demonstrate that it is resilient against poorly-edited target images and propose a simple extension to predict image sequences with a strict time budget of 1-2 seconds per image.
\end{abstract} 






%\allowdisplaybreaks
%\raggedbottom

\begin{teaserfigure}
  \centering
  \includegraphics[width=1.0\linewidth]{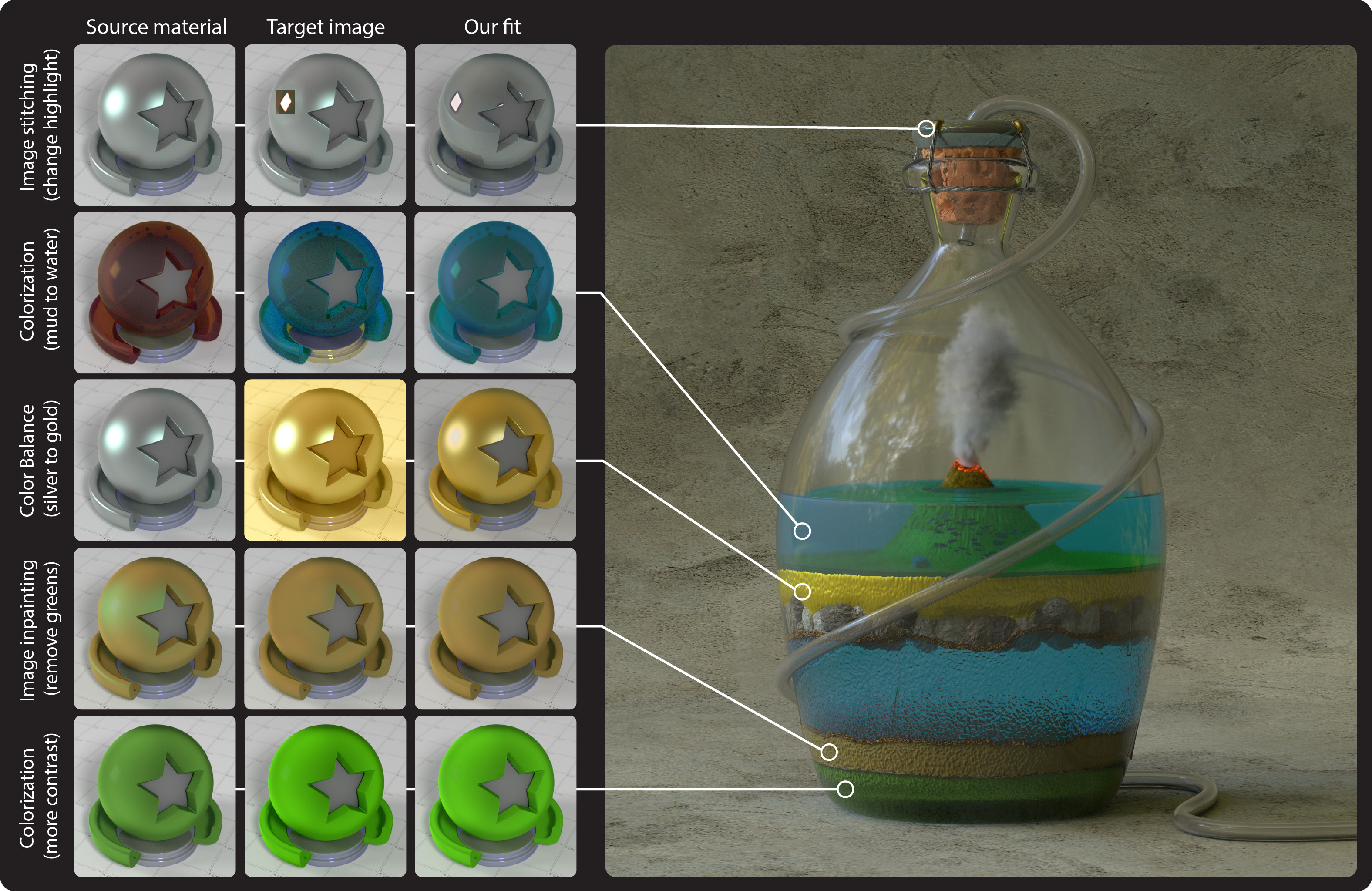}
  %\vspace{.4cm}
  % Our technique enables users without expertise in photorealistic material editing to create high-quality materials: in our workflow, the artist starts out with a source material, edits it directly using standard image processing techniques,
  \caption{We propose a hybrid technique to empower novice users and artists without expertise in photorealistic rendering to create sophisticated material models by applying standard image editing operations to a source image. Then, in the next step, our method proceeds to find a photorealistic BSDF that, when rendered, resembles this target image. Our method generates each of the showcased fits within 20-30 seconds of computation time and is able to offer high-quality results even in the presence of poorly-executed edits (e.g., the background of the gold target image, the gold-colored pedestal for the water material and the stitched specular highlight above it). \Revision{Scene: Reynante Martinez.}}
  \label{fig:teaser}
\end{teaserfigure}

\maketitle

\begin{figure*}[h!]
\includegraphics[width=0.93\textwidth]{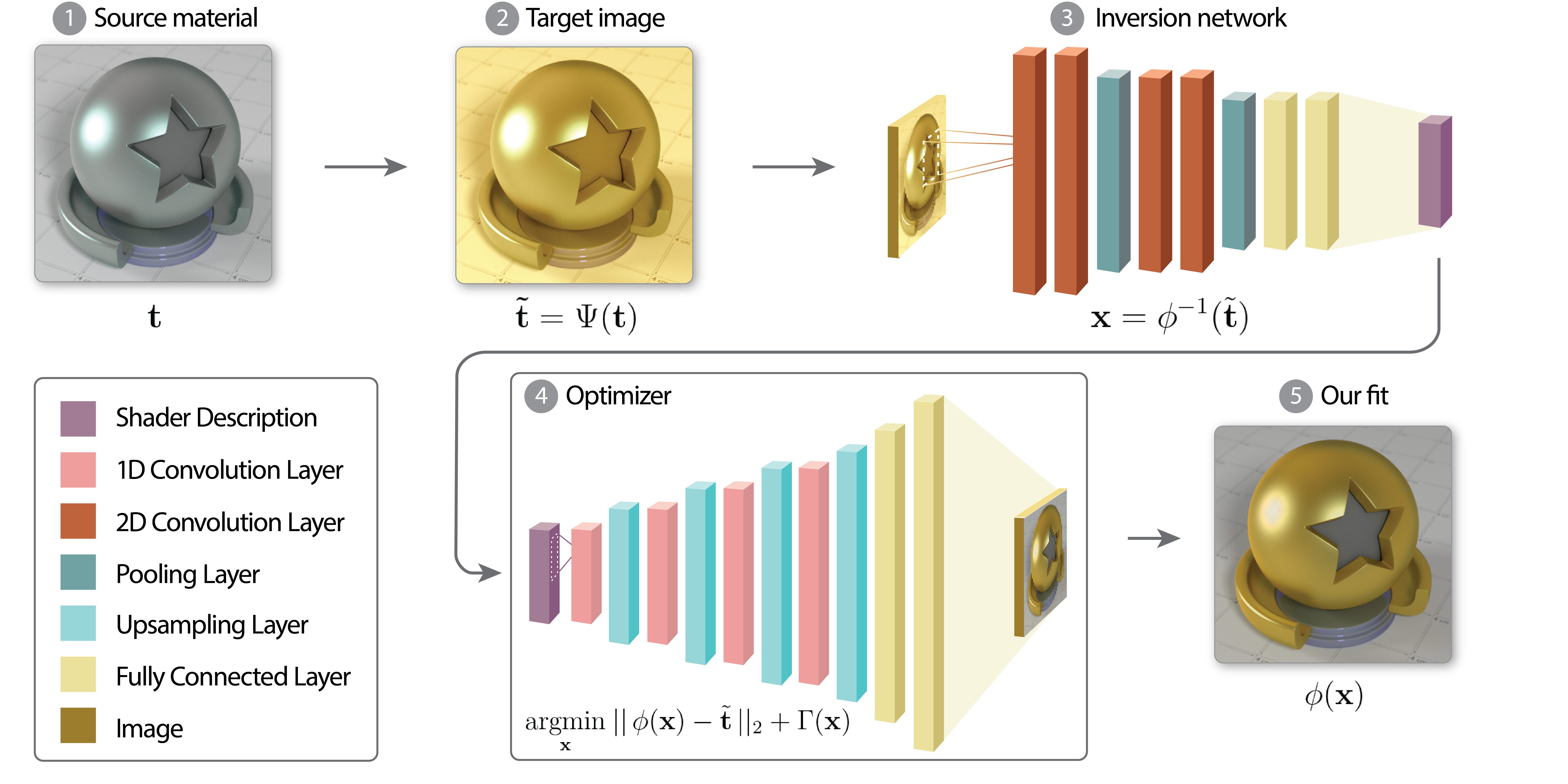}
\caption{Our proposed hybrid technique offers an intuitive workflow where the artist takes a source material (\ding[1.15]{182}) and produces the target image by applying the desired edits to it within a 2D raster image editor of their choice (\ding[1.15]{183}). Then, one or more encoder neural networks are used to propose a set of approximate initial guesses (\ding[1.15]{184}) to be used with our neural network-augmented optimizer (\ding[1.15]{185}), which rapidly finds a photorealistic shader setup that closely matches the target image (\ding[1.15]{186}). The artist then finishes the process by assigning this material to a target object and renders the final scene offline.}
\label{fig:workflow}
\end{figure*}

%\includegraphics[width=1.0\linewidth]{figures/teaser3.png}
%  \centering
%   \caption{
%	Our technique opens up the possibility of creating photorealistic materials by using general image processing operators: the artist starts out with a base material, applies non-physical changes to it, and our method finds the closest possible photorealistic BRDF to match it. %These materials can then be used to populate a scene.
   %Our proposed method generates each of the showcased fits within 20 seconds of computation time and was able to offer high-quality results even in the presence of poorly executed edits (e.g., the background of the gold image and gold-colored pedestal in the target image for the water material).
% }
% \label{fig:teaser}

% The visual difference between photorealistic rendering and rasterization techniques has shrunk significantly over the last decade. 

\begin{figure}[ht!]
\includegraphics[width=0.49\textwidth]{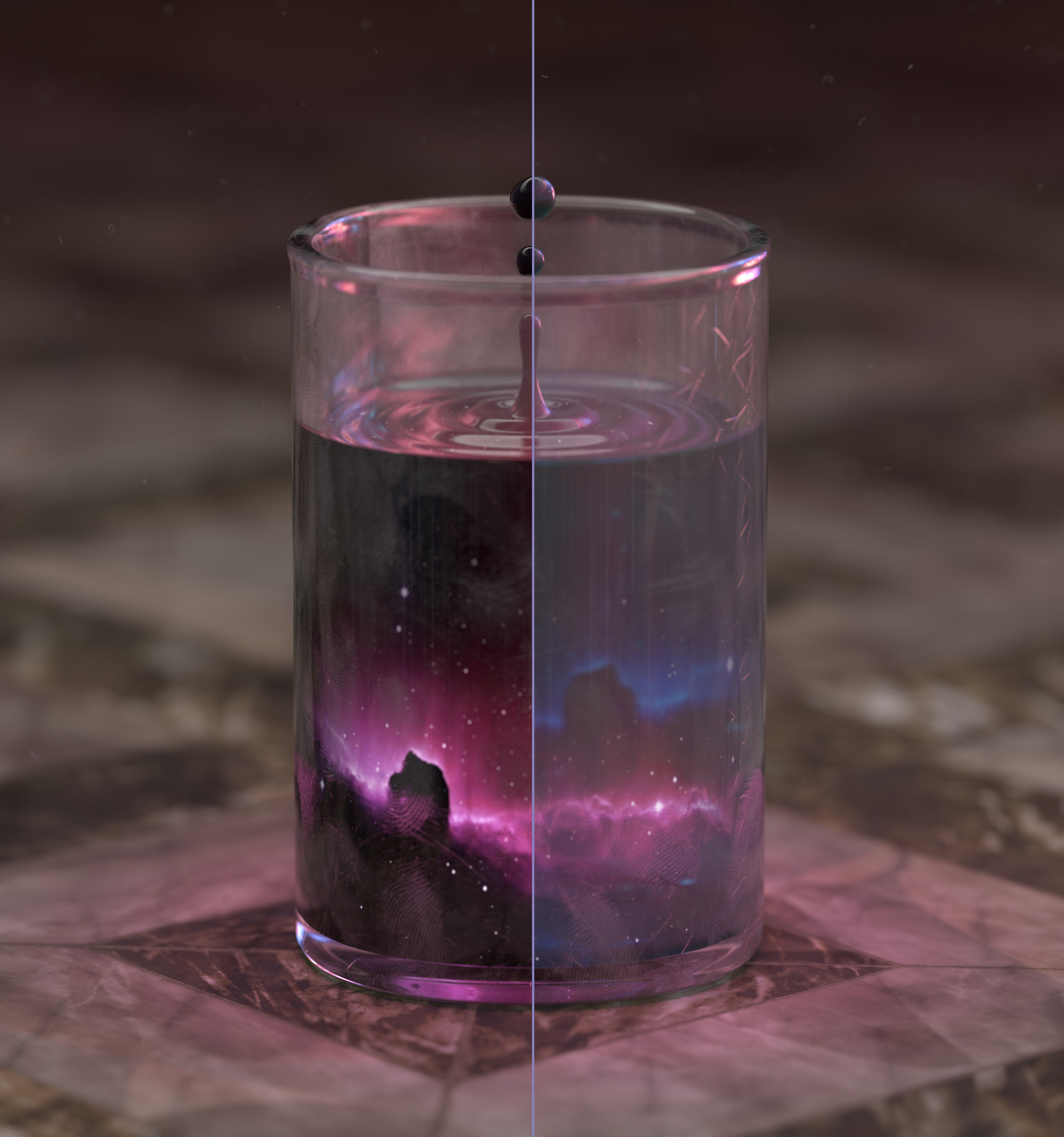}
\caption{To demonstrate the utility of our system, we synthesized a new material and deployed it into an already existing scene using Blender and Cycles. In this scene, we made a material mixture to achieve a richer and foggier nebula effect inside the glass. Left: theirs, right: 50\% theirs, 50\% ours. \Revision{Scene: Reynante Martinez.}}
% neural opt, nn, hybrid, target
\label{fig:glass-scene}
\end{figure}

\section{Introduction}
The expressiveness of photorealistic rendering systems has seen great strides as more sophisticated material models became available for artists to harness. Most modern rendering systems offer a node-based shader tool where the user can connect different kinds of material models and perform arbitrary mathematical operations over them (e.g., addition and mixing), opening up the possibility of building a richer node graph that combines many of the more rudimentary materials to achieve a remarkably expressive model. These are often referred to as ``principled'' shaders and are commonly used within the motion picture industry \cite{burley2012physically}. However, this expressiveness comes with the burden of complexity, i.e., the user has to understand each of the many parameters of the shader not only in isolation, but also how they influence each other, which typically requires years of expertise in photorealistic material modeling. In this work, we intend to provide a tool that can be used by a wider target audience, i.e., artists and novices that do not have any experience creating material models, but are adept at general-purpose image processing and editing. This is highly desirable as human thinking is inherently visual and is not based on physically-based material parameters \cite{roder2002speech,white1989visual}. We propose a workflow in which the artist starts out with an image of a reference material and applies classic image processing operations to it. Our key observation is that even though this processed target image is often not physically achievable, in many cases, a photorealistic material model can be found that is remarkably close to it (Fig. \ref{fig:workflow}). These material models can then be easily inserted into already existing scenes by the user (Fig. \ref{fig:glass-scene}).

In summary, we present the following contributions:
\begin{itemize}
\item An optimizer that can rapidly match the target image when given an approximate initial guess.
\item A neural network to solve the adjoint rendering problem, i.e., take the target image as an input and infer a shader that produces a material model to approximate it.
\item A hybrid method that combines the advantages of these two concepts and achieves high-quality results for a variety of cases within 30 seconds.
\item A simple extension of our method to enable predicting sequences of images within 1-2 seconds per image.
\end{itemize}
%\emph{Upon publishing this work, we will provide our pre-trained neural networks and the source code for the entirety of this project.}
\href{https://users.cg.tuwien.ac.at/zsolnai/gfx/photorealistic-material-editing/}{\textcolor{blue}{We provide our pre-trained neural network and the source code for the entirety of this project here: \url{https://users.cg.tuwien.ac.at/zsolnai/gfx/photorealistic-material-editing/}}}
%\Revision{We provide our pre-trained neural network and the source code for the entirety of this project.}%\footnote{Link: \url{https://users.cg.tuwien.ac.at/zsolnai/gfx/gaussian-material-synthesis/}}}
%We provide our pre-trained neural network and the source code for the entirety of this project.
%We provide our pre-trained neural network and the source code for the entirety of this project under the following link: \newline \scriptsize{\textcolor{blue}{https://users.cg.tuwien.ac.at/zsolnai/gfx/gaussian-material-synthesis/}}

%\begin{figure*}[ht]
%\includegraphics[width=1.0\textwidth]{figures/workflow.jpg}
%\caption{Our technique opens up the possibility of creating photorealistic materials by using general image processing operators: the artist starts out with a base material, applies non-physical changes to it, and our method finds the closest possible photorealistic BRDF to match it. These materials can then be used to populate a scene.}
%\label{fig:workflow}
%\end{figure*}

%but often lack support for several important aspects of light transport, e.g., refractive and translucent materials (i.e., BSSRDFs\footnote{Bidirectional scattering-surface reflectance distribution functions.})

\section{Previous Work}
\subsection{Material Acquisition}
A common workflow for photorealistic material acquisition requires placing the subject material within a studio setup and using measurement devices to obtain its reflectance properties \cite{marschner1999image,miyashita2016zoematrope}. To import this measured data into a production renderer, it can be either used as-is, can be compressed down into a lower-dimensional representation \cite{papas2013fabricating,Rainer2019Neural} or approximated through an analytic BSDF\footnote{Bidirectional scattering distribution function.} model \cite{papas2014physically}. Many recent endeavors improve the cost efficiency and convenience of this acquisition step by only requiring photographs of the target material \cite{aittala2015two,aittala2016reflectance,deschaintre2018single,li2017modeling,li2018materials} while still requiring physical access to these source material samples, while precomputed BSDF databases offer an enticing alternative where the user can choose from a selection of materials \cite{matusik2003data,Dupuy2018Adaptive}. We aim to provide a novel way to exert direct artistic control over these material models.
Our method can be related to inverse rendering approaches \cite{marschner1998inverse,ramamoorthi2001signal}, where important physical material properties are inferred from a real photograph with unknown lighting conditions. In our work, the material test scene contains a known lighting and geometry setup, but in return, enables not only the rapid discovery of new materials, but artistic control through standard and well-known image-space editing operations.

\subsection{Material Editing}
To be able to efficiently use the most common photorealistic rendering systems, an artist is typically required to have an understanding of physical quantities pertaining to the most commonly modeled phenomena in light transport, e.g., indices of refraction, scattering and absorption albedos and more \cite{song2009subedit,burley2012physically}. This modeling time can be cut down by techniques that enable editing BRDF\footnote{Bidirectional reflectance distribution function.} models directly within the scene \cite{ben2006real,cheslack2008fast,sun2007interactive}, however, with many of these methods, the artist is still required to understand the physical properties of light transport, often incurring a significant amount of trial and error. Instead of editing the materials directly, other techniques enable editing secondary effects, such as caustics and indirect illumination within the output image \cite{schmidt13,ben2008precomputed}. \Revision{An efficient material editing workflow also opens up the possibility of rapid relighting previously rendered scenes \cite{wang2008real,ng2004triple,wang2004all}.} Reducing the expertise required for material editing workflows has been a subject to a large volume of research works: an intuitive editor was proposed by pre-computing many solutions to enable rapid exploration \cite{havsan2013interactive}, carefully crafted material spaces were derived to aid the artist \cite{serrano2016intuitive,soler2018versatile,lagunas2019similarity}, and learning algorithms have been proposed to create a latent space that adapts to the preferences of the user \cite{Zsolnai18}. We endeavored to create a solution that produces the desired results \emph{rapidly} by looking at a non-physical mockup image, requiring expertise only in 2D image editing, which is considered to be common knowledge by nearly all artists in the field. \Revision{Generally, BRDF relighting methods are preferable when in-scene editing is a requirement, otherwise, we recommend using our proposed technique in the case of one sought material to moderate-scale problems and Gaussian Material Synthesis (GMS) for mass-scale material synthesis.}

\Revision{
\subsection{Neural Networks and Optimization}
%The study of using optimization techniques along with neural networks has been subject to a significant amount of research. 
Optimization is present at the very core of every modern neural network: to be able to minimize the prescribed loss function efficiently, the weights of the networks are fine-tuned through gradient descent variants \cite{bottou2010large,robbins1951stochastic} or advanced methods that include the use of lower-order moments \cite{kingma2014adam}, while additional measures are often taken to speed up convergence and avoid poor local minima \cite{sutskever2013importance,goh2017momentum}. Similar optimization techniques are also used to generate the model description and architecture of these neural networks \cite{zoph2016neural,elsken2018neural}, or the problem statement itself can also be turned around by using learning-based methods to discover novel optimization methods \cite{bello2017neural}. In this work, we propose two combinations of a neural network and an optimizer -- first, the two can be combined \emph{indirectly} by endowing the optimizer with a reasonable initial guess, and \emph{directly} by using the optimizer that invokes a neural renderer at every function evaluation step to speed up the convergence by several orders of magnitude (steps \ding[1.15]{184} and \ding[1.15]{185} in Fig. \ref{fig:workflow}). This results in an efficient two-stage system that is able to rapidly match a non-physical target image and does not require the user to stay within a prescribed manifold of artistic editing operations \cite{zhu2016generative}.
}

%  or changing the input image itself \cite{davisimage}.

\section{Overview}
\label{sec:overview}

Many trained artists are adept at creating new photorealistic materials by engaging in a direct interaction with a principled shader. This workflow includes adjusting the parameters of this shader and waiting for a new image to be rendered that showcases the appropriate output material. If at most a handful of materials are sought, this is a reasonably efficient workflow, however, it also incurs a significant amount of rendering time and expertise in material modeling. Our goal is to empower novice and intermediate-level users to be able to reuse their knowledge from image processing and graphic design to create their envisioned photorealistic materials.

In this work, we set up a material test scene that contains a known lighting and geometry setup, and a fixed principled shader with a vector input of $x \in \mathds{R}^m$ where $m=19$. This shader is able to represent the most commonly used diffuse, glossy, specular and translucent materials with varying roughness and volumetric absorption coefficients. Each parameter setup of this shader produces a different material model when rendered. In our workflow, the user is offered a variety of images, and chooses one desired material model as a starting point. Then, the user is free to apply a variety of image processing operations on it, e.g., colorization, image inpainting, blurring a subset of the image and more. Since these image processing steps are not grounded in a physically-based framework, the resulting image is not achievable by adjusting the parameters in the vast majority of cases. However, we show that our proposed method is often able a produce a photorealistic material that closely matches this target image.

\textbf{Solution by optimization.} When given an input image $\myvec{t} \in \mathds{R}^p$, it undergoes a series of transformations (e.g., colorization, image inpainting) as the artist produces the target image $\tilde{\myvec{t}}=\Psi(\myvec{t})$, where $\Psi: \mathds{R}^p \rightarrow \mathds{R}^p$. Then, an image is created from an initial shader configuration, i.e., $\phi\!: \mathds{R}^m \!\! \rightarrow \! \mathds{R}^p$, where $m$ refers to the number of parameters within the shader and $p$ is the number of variables that describe the output image (in our case $p=3\cdot410^2$ is used with the range of 0-255 for each individual pixel). This operation is typically implemented by a global illumination renderer. Our goal is to find an appropriate parameter setup of the principled shader $\myvec{x} \in \, \mathds{R}^m$ that, when rendered, reproduces $\tilde{\myvec{t}}$. Generally, this is not possible as a typical $\Psi$ leads to images that cannot be perfectly matched through photorealistic rendering. However, surprisingly, we can often find a configuration $\myvec{x}$ that produces an image that closely resembles $\tilde{\myvec{t}}$ through solving the minimization problem
\begin{align*}
	\argmin_{\myvec{x}} \,\,\,\,\,\,\,\, \,\,\,\, &|| \, \phi(\myvec{x})  - \tilde{\myvec{t}} \, ||_{2}, \\
	\text{subject to} \,\,\,\,\,\,\,\, \,\,\,\, &\myvec{x}_{\text{min}} \leq \myvec{x} \leq \myvec{x}_{\text{max}}, \numberthis
	\label{eq:optimization}
\end{align*}
where the constraints stipulate that each shader parameter has to reside within the appropriate boundaries (i.e., $0 \leq x_i \leq 1$ for albedos or $x_j \geq 1$ for indices of refraction \Revision{where $x_i, x_j \! \in \! \myvec{x}$}). To be able to benchmark a large selection of optimizers, we introduce an equivalent alternative formulation of this problem where the constraints are reintroduced as a barrier function $\Gamma(\cdot)$, i.e., 
\begin{align*}
	\argmin_{\myvec{x}} &\,\,\,\, \Big( || \, \phi(\myvec{x})  - \tilde{\myvec{t}} \, ||_{2} + \Gamma(\myvec{x}) \Big), \text{   where}\\
	\Gamma(\myvec{x}) &=
	\begin{dcases}
		0, & \text{if $\myvec{x} \in \mathcal{C}$}, \\
		+\infty, & \text{otherwise},
	\end{dcases} \\
	\mathcal{C} \,\,\,\, &= \, \Big\{  \myvec{x} \,\,\, | \,\,\, f_i(\myvec{x}) \geq \myvec{0}, \,\, i=1,2 \Big\}, \\
	f_1(\myvec{x}) &= \myvec{x}_{\text{max}} - \myvec{x}, \\
	f_2(\myvec{x}) &= \myvec{x} - \myvec{x}_{\text{min}}. \label{eq:our_optimization} \numberthis \\
\end{align*}
where $\mathcal{C}$ denotes the feasible region chosen by a set of constraints described by $f_i(\cdot)$ (equivalent to the second line in (\ref{eq:optimization})) and the vector comparison operator ($\geq$) here is considered true only when all of the vector elements exceed (or equal to) zero. In a practical implementation, the infinity can be substituted by a sufficiently large integer. This formulation enabled us to compare several optimizers (Table \ref{tab:opt-test} in Appendix \ref{sec:appB}), where we found Nelder and Mead's simplex-based self-adapting optimizer \shortcite{nelder1965simplex} to be the overall best choice due to its ability to avoid many poor local minima through its contraction operator and used that for each of the reported results throughout this manuscript.

\begin{figure*}[ht!]
\includegraphics[width=0.95\textwidth]{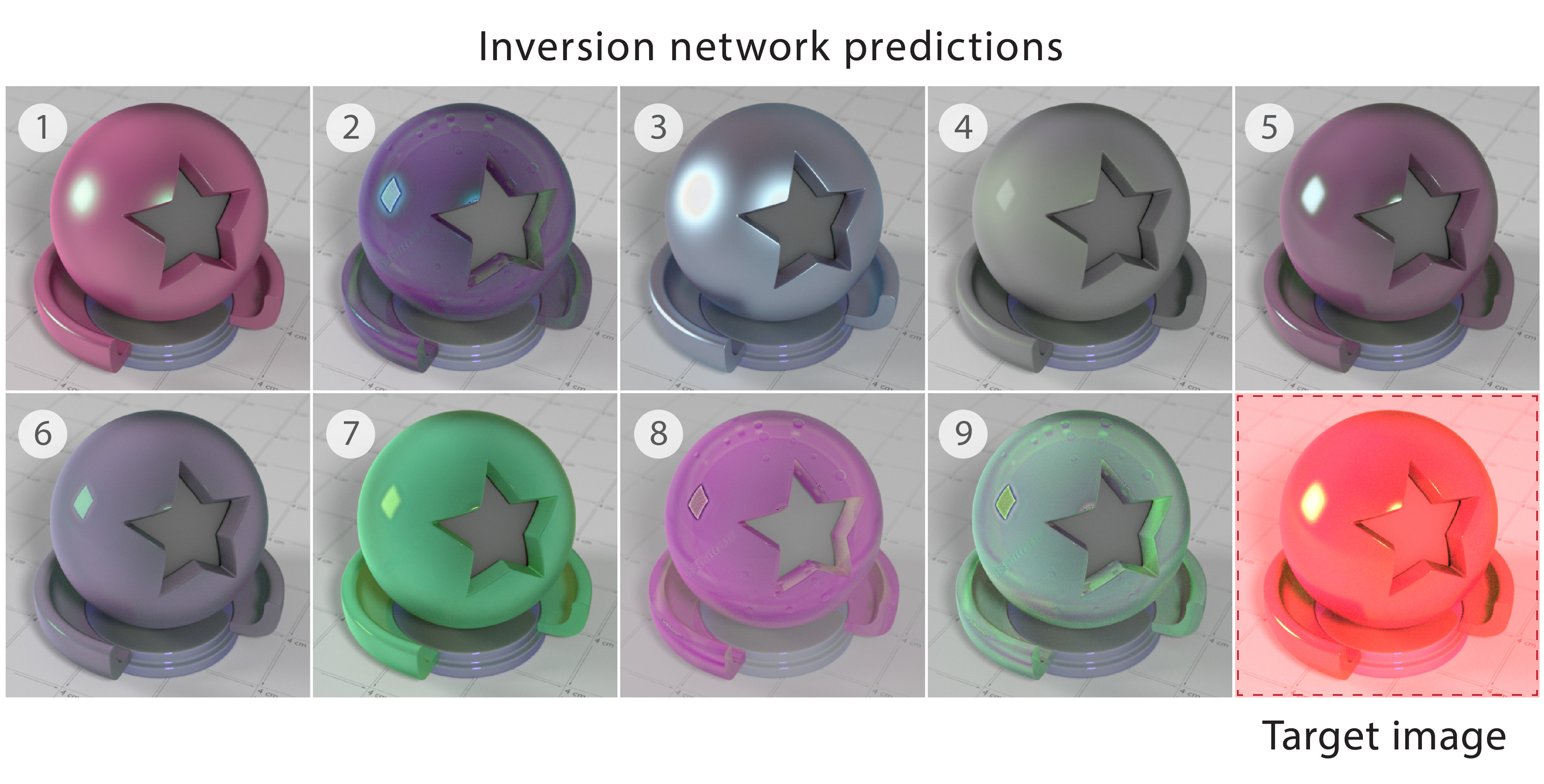}
\caption{Whenever the target image strays too far away from the images contained within their training set (lower right), our 9 inversion networks typically fail to provide an adequate solution. However, using our ``best of n'' scheme and our hybrid method, the best performing prediction of our neural networks can be used to equip our optimizer with an initial guess, substantially improving its results.}
\label{fig:nn_fail}
\end{figure*}

%where $\mathcal{C}$ denotes the feasible region chosen by a set of constraints described by $f_i(\cdot)$ (equivalent to the last line in (\ref{eq:optimization})). We also used $\myvec{x}$ instead of $\myvec{x}$ to indicate that this is a surrogate problem that deviates from the previous, constrained formulation. If we solve this unconstrained variant several times with increasing $k$ and reusing the optima found in previous steps as initial guesses, as $k \to +\infty$, $\myvec{x}_{\text{opt}}^{*} \to \myvec{x}_{\text{opt}}$ also holds, i.e., we obtain the solution to the original, constrained problem described in (\ref{eq:optimization}). As $\Gamma(\cdot)$ is smooth and convex within $\mathcal{C}$ and tends towards infinity as $(\myvec{x})$ approaches the boundary, this barrier function can be intuitively interpreted as a force field that repels a particle to remain within $\mathcal{C}$. 
% \footnote{In a practical implementation, the infinity can be substituted by a sufficiently large integer. to avoid singularities during the multiplication with $\epsilon$.}.

 Nonetheless, solving this optimization step still takes several hours as each function evaluation invokes $\phi$, i.e., a rendering step to produce an image, which clearly takes too long for day-to-day use in the industry. We introduce two solutions to remedy this limitation, followed by a hybrid method that combines their advantages. \newline \newline
\textbf{Neural renderer.} 
To speed up the function evaluation process, we replace the global illumination engine that implements $\phi$ with a neural renderer \cite{Zsolnai18}. This way, instead of running a photorealistic rendering program at each step, our optimizer invokes the neural network to predict this image, thus reducing the execution time of the process by several orders of magnitude, in our case, from an average of 50 seconds to 4ms per image at the cost of restricting the material editing to a prescribed scene and lighting setup. Because of the lack of a useful initial guess, this solution still requires many function evaluations and is unable to reliably provide satisfactory solutions. \newline \newline
\textbf{Solution by inversion.}
One of our key observations is that an approximate solution can also be produced \emph{without} an optimization step by finding an appropriate inverse to $\phi$: since $\phi$ is realized through a decoder neural network (i.e., neural renderer) that produces an image from a shader configuration, $\phi^{-1}$, its inverse, can be implemented as an \emph{encoder} network that takes an image as an input and predicts the appropriate shader parameter setup that generates this image. This adjoint problem has several advantages: first, such a neural network can be trained on the same dataset as $\phi$ by only swapping the inputs and outputs and retains the advantageous properties of this dataset, e.g., arbitrarily many new training samples can be generated via rendering, thereby loosening the ever-present requirement of preventing overfitting via regularization \cite{srivastava2014dropout,nowlan1992simplifying,zou2005regularization}. Second, we can use it to find a solution \emph{directly} through $\myvec{x} \approx \phi^{-1}(\tilde{\myvec{t}})$ without performing the optimization step described in (\ref{eq:optimization}-\ref{eq:our_optimization}). As the output image is not produced through a lengthy optimization step, but is inferred by this encoder network, this computes in a few milliseconds. We will refer to this solution as the \emph{inversion network} and note that our implementation of $\phi^{-1}$ only approximately admits the mathematical properties of a true inverse function. We also discuss the nature of the differences in more detail in Section \ref{sec:results}. We have trained 9 different inversion network architectures and found that typically, each of them performs well on a disjoint set of inputs. Our other key observation is that because we have an atypical problem where the ground truth image ($\tilde{\myvec{t}}$) is available and each of the candidate images can be inferred inexpensively (typically within 5 milliseconds), it is possible to compute a ``best of $n$'' solution by comparing all of these predictions to the ground truth, i.e.,
%However, this result may be inaccurate if the transformed target image strays far from the inputs contained within the training set (Fig. \ref{fig:nn_fail}). 
\begin{align*}
%&\myvec{x} = \phi_{(i)}^{-1}(\tilde{\myvec{t}}), \text{ where} \\ 
%i = \argmin_{i}& \big\{ || \phi(\phi_{(i)}^{-1}(\tilde{\myvec{t}})) - \tilde{\myvec{t}} \, ||_{2} \big\}, \, i=(1,\ldots,n), \numberthis
\myvec{x} = \phi_{(i)}^{-1}(\tilde{\myvec{t}}), \,\, \text{where} \,\,\,\,\, i = \argmin_{j}& \,\, || \phi(\phi_{(j)}^{-1}(\tilde{\myvec{t}})) - \tilde{\myvec{t}} \, ||_{2}, \numberthis
\label{eq:bestof9}
\end{align*}
where $\phi_{(i)}^{-1}$ denotes the prediction of the $i$-th inversion network, $j=(1,\ldots,n)$, and in our case, $n\!=\!9$ was used. This step introduces a negligible execution time increase and in return, drastically improves the quality of this inversion process for a variety of test cases. However, these solutions are only approximate in cases where the target image strays too far away from the training data (Fig. \ref{fig:nn_fail}). In Appendix \ref{sec:appA} we report the structure of the neural networks used in this figure. \newline \newline
\textbf{Hybrid solution.}
Both of our previous solutions suffer from drawbacks: the optimization approach provides results that resemble $\tilde{\myvec{t}}$ but is impracticable due to the fact that it requires too many function evaluations and gets stuck in local minima, whereas the inversion networks rapidly produce a solution, but offer no guarantees when the target image significantly differs from the ones shown in the training set. We propose a hybrid solution based on the knowledge that even though the inverse approach does not provide a perfect solution, since it can produce results instantaneously that are significantly closer to the optimum than a random input, it can be used to endow the optimizer with a reasonable initial guess. This method is introduced as a variant of (\ref{eq:our_optimization}) where $\myvec{x}_{\text{init}} = \phi^{-1}(\tilde{\myvec{t}})$ and a more detailed description of this hybrid solution is given below in Algorithm \ref{alg:pme}. Additionally, this technique is able to not only provide a ``headstart'' over the standard optimization approach but was also able to find higher quality solutions in all of our test cases. \newline \newline

\begin{figure}[htb!]
\includegraphics[width=0.49\textwidth]{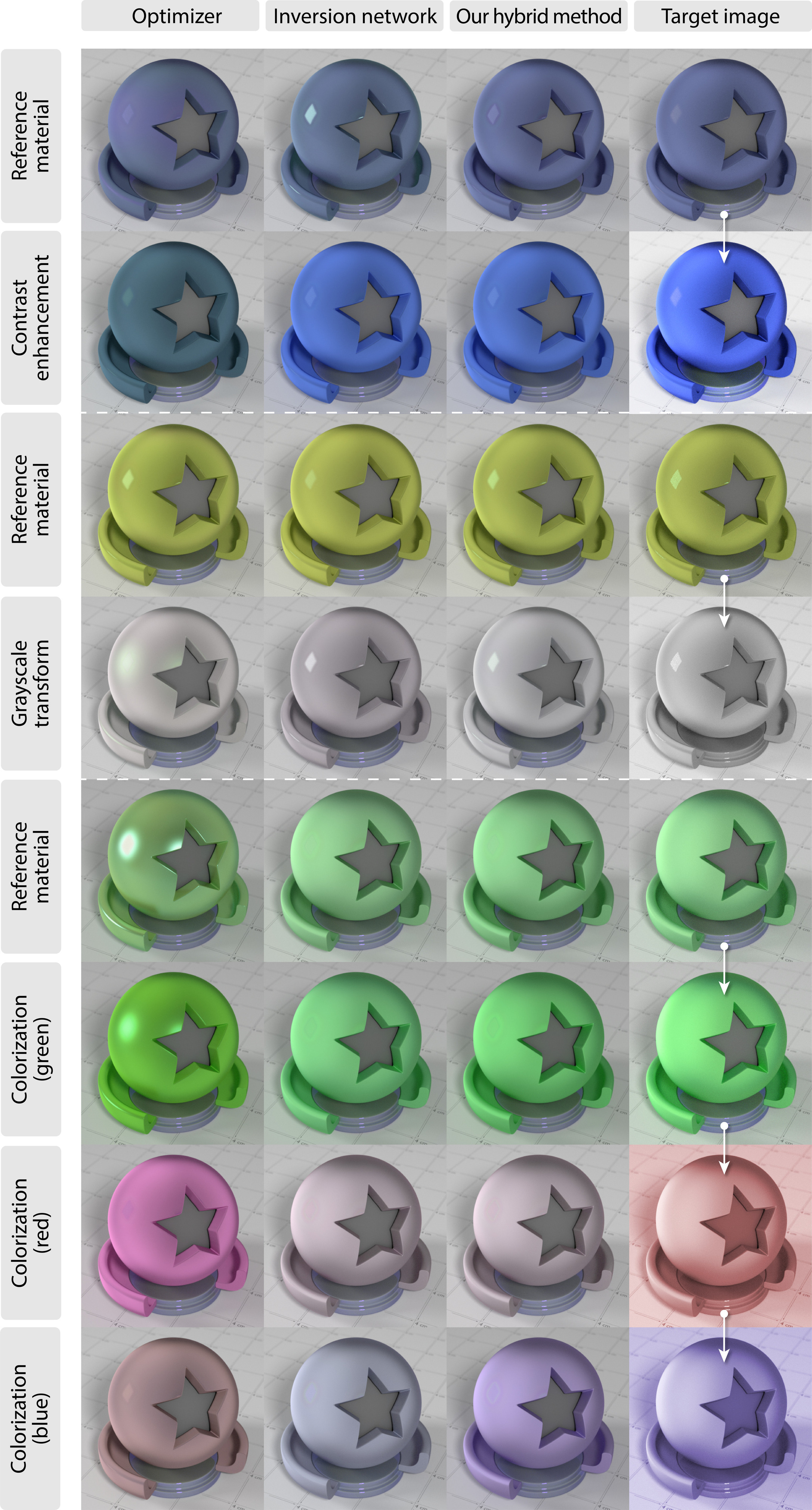}
\caption{Results for three techniques on common global colorization operations including saturation increase and grayscale transform. The ``reference material'' labels showcase materials that can be obtained using our shader and are used as source images for the materials below them, where the arrows denote the evolution of the target image.}
% neural opt, nn, hybrid, target
\label{fig:global1}
\end{figure}

\begin{algorithm}
  \caption{Photorealistic Material Editing
    \label{alg:pme}}
  \begin{algorithmic}[1]
    %\Statex
	\State \textbf{Given} $\myvec{t}, \, \phi(\cdot), \,\,\, \big[\phi_{(1)}^{-1}(\cdot), \ldots, \phi_{(n)}^{-1}(\cdot)\big], \,\,\, \myvec{x}_{\text{min}}, \, \myvec{x}_{\text{max}}$
	%\State \Comment{Input image, neural renderer, image predictor networks, parameter constraints}
	\State $\tilde{\myvec{t}} \gets \Psi(\myvec{t})$ \Comment{Obtain target image}
	\For{$i \gets 1 \textrm{ to } n$} \Comment{Predict with $n$ inversion networks}
		\State Compute each $\phi_{(i)}^{-1}(\tilde{\myvec{t}})$
	\EndFor
	\State \textbf{Find} i = $\argmin_{j \in 1..n} || \phi(\phi_{(j)}^{-1}(\tilde{\myvec{t}})) - \tilde{\myvec{t}} \, ||_{2} $ \Comment{Find best candidate}
	\State Define $\myvec{x}_{\text{init}} \gets \phi_{(i)}^{-1}(\tilde{\myvec{t}})$
	\State Define $f_1(\myvec{x}) = \myvec{x}_{\text{max}} - \myvec{x}$ \Comment{Set up constraints}
	\State Define $f_2(\myvec{x}) = \myvec{x} - \myvec{x}_{\text{min}}$
	\State Define $\mathcal{C} = \, \big\{  \myvec{x} \,\,\, | \,\,\, f_i(\myvec{x}) \geq \myvec{0}, \,\, i=1,2 \big\}$ \Comment{Construct feasible region}
	\State Define $\Gamma(\myvec{x}) =
	\begin{dcases}
		0, & \text{if $\myvec{x} \in \mathcal{C}$}, \\
		+\infty, & \text{otherwise}
	\end{dcases}$ \Comment{Construct barrier}
	\State \textbf{Initialize} optimizer with $\myvec{x}_{\text{init}}$
	\State \textbf{Minimize} $\argmin_{\myvec{x}} \, \big( || \phi(\myvec{x})  - \tilde{\myvec{t}} ||_{2} + \Gamma(\myvec{x}) \big)$ \Comment{Refine initial guess}
	\State Display $\phi(\myvec{x})$ to user
  \end{algorithmic}
\end{algorithm}
%\begin{equation}
	%\Psi_{\sigma}(\myvec{t}) = G(x,y,\sigma) \, \myvec{t} = \frac{\myvec{t}}{2\pi\sigma^2} \exp\Big(-\frac{x^2+y^2}{2\sigma^2}\Big),
%\end{equation}
%\begin{equation}
 	%\tilde{\myvec{t}}_{(k)} = \Psi_{k}(\myvec{t}),
 	%\tilde{\myvec{t}}_{(0)} &= \myvec{t} \numberthis \\
%\end{equation}
%In the interest of readability, we have omitted a conversion step between the flattened 1D and regular 2D representations of $\myvec{t}$ and 
\textbf{Predicting image sequences.}
A typical image editing workflow takes place within a raster graphics editor program where the artist endeavors to find an optimal set of parameters,  e.g., the kernel width $\sigma$ in the case of a Gaussian blur operation to obtain their envisioned artistic effect. This process includes a non-trivial amount of trial and error where the artist decides whether the parameters should be increased or decreased; this is only possible in the presence of near-instant visual feedback that reflects the effect of the parameter changes on the image. We propose a simple extension to our hybrid method to accommodate these workflows: consider an example scenario where the $k$-th target image in a series of target images $\tilde{\myvec{t}}_{(k)}$ are produced by subjecting a starting image $\myvec{t}$ to an increasingly wide blurring kernel. This operation is denoted by $\Psi_{\sigma}(\mathbf{t}) = G_{\sigma} * \mathbf{t}$, where $G_{\sigma}$ is a zero-centered Gaussian, and for simplicity, the target images are produced via $\tilde{\myvec{t}}_{(k)} = \Psi_{k}(\myvec{t})$, with the initial condition of $\tilde{\myvec{t}}_{(0)}\!=\!\myvec{t}$. We note that many other transforms can also be substituted in the place of $\Psi$ without loss of generality. We observe that such workflows create a series of images where each neighboring image pair shows only minute differences, i.e., for any positive non-zero $k$, $||\tilde{\myvec{t}}_{(k+1)} - \tilde{\myvec{t}}_{(k)}||_{2}$ remains small. As in these cases, we are required to propose many output images, we can take advantage of this favorable mathematical property by extending the pool of initial inversion networks with the optimized result of the previous frame by modifying Steps 3-5 of Algorithm \ref{alg:pme} to add
\begin{equation}
\phi_{(n+1)}^{-1}(\tilde{\myvec{t}}_{k}) = \argmin_{\myvec{x}} \, \Big( || \phi(\myvec{x}) - \tilde{\myvec{t}}_{k-1} ||_{2} + \Gamma(\myvec{x}) \Big).
\end{equation}
Note that this does not require any extra computation as the result of Step 12 of the previous run can be stored and reused. Intuitively, this means that \emph{both} the inversion network predictions and the prediction of the previous image are used as candidates for the optimization (whichever is better). This way, after the optimization step is finished, the improvements can be ``carried over'' to the next frame. This method we refer to as \emph{reinitialization} and in Section \ref{sec:results}, we show that it consistently improves the quality of our output images for such image sequences, even with a strict budget of 1-2 seconds per image. \newline \newline
%\Revision{
%Even in the presence of several inversion networks, the output results may be inaccurate if the transformed target image strays far from the inputs contained within the training set.
%}

%\begin{table}[t]
%\centering
%	\begin{tabular}{cccccc}
%		  \toprule
%		  Input & Neural opt. & Neural net. & Hybrid \\
%		  \midrule
%		  Fig. \ref{fig:global1}, Row 2 & 4.99 (34.72 / 11.57) & 8.87 & $\mathbf{3.90}$  \\
%          Fig. \ref{fig:global1}, Row 4 & 6.34 (24.74 / 10.48) & 12.57 & $\mathbf{6.21}$  \\
%          Fig. \ref{fig:global1}, Row 7 & 13.40 (39.90 / 19.75) & 28.80 & $\mathbf{13.19}$ \\
%          Fig. \ref{fig:global1}, Row 8 & 3.82 (34.23 / 12.31) & 20.67 & $\mathbf{3.72}$ \\
%          Fig. \ref{fig:local1}, Row 2 & 12.24 (49.63/19.67) & 41.51 & $\mathbf{12.07}$ \\
%          Fig. \ref{fig:local1}, Row 3 & 12.24 (49.63/19.67) & 41.51 & $\mathbf{12.07}$ \\
%          Fig. \ref{fig:local1}, Row 8 & 12.24 (49.63/19.67) & 41.51 & $\mathbf{12.07}$ \\
          %\cdashlinelr{1-6}
%		  \bottomrule
%	\end{tabular}
%	\caption{A comparison of our proposed techniques. To be recomputed later. Ignore this table now.}
%\label{tab:rprop}
%\end{table}

\begin{table*}[ht]
\centering
  \begin{tabular}{ccccccccccc}
      \toprule
      & \multicolumn{2}{c}{Initial guess} & \multicolumn{2}{c}{50 fun. evals} & \multicolumn{2}{c}{300 fun. evals} & \multicolumn{2}{c}{1500 fun. evals} \\
      \cmidrule(lr){2-3}
      \cmidrule(lr){4-5}
      \cmidrule(lr){6-7}
      \cmidrule(lr){8-9}
      Input & Random & NN & Optimizer & Ours & Optimizer & Ours & Optimizer & Ours \\
      \midrule
          Fig. \ref{fig:global1}, Row 1 & 41.93 & 5.94 & 33.81 & $\mathbf{4.53}$ & 9.42 & $\mathbf{2.84}$ & 5.62 & $\mathbf{2.37}$ \\
          \cdashlinelr{1-9}
          Fig. \ref{fig:global1}, Row 2 & 78.45 & 32.72 & 68.55 & $\mathbf{32.67}$ & 40.24 & $\mathbf{32.67}$ & 40.21 & $\mathbf{32.67}$ \\
          Fig. \ref{fig:global1}, Row 4 & 35.37 & 18.68 & 30.88 & $\mathbf{16.53}$ & 17.29 & $\mathbf{14.71}$ & 16.98 & $\mathbf{14.68}$ \\ 
          Fig. \ref{fig:global1}, Row 7 & 41.65 & 22.42 & 38.10 & $\mathbf{22.38}$ & 26.30 & $\mathbf{22.38}$ & 26.24 & $\mathbf{22.38}$ \\
          Fig. \ref{fig:global1}, Row 8 & 29.04 & 19.82 & 26.79 & $\mathbf{18.43}$ & 22.93 & $\mathbf{15.37}$ & 22.93 & $\mathbf{15.37}$ \\
          Fig. \ref{fig:local1}, Row 2 & 23.78 & 12.79 & 20.31 & $\mathbf{11.62}$ & 8.27 & $\mathbf{7.81}$ & 8.26 & $\mathbf{7.80}$ \\
          Fig. \ref{fig:local1}, Row 3 & 21.60 & 9.09 & 16.54 & $\mathbf{8.28}$ & 6.24 & $\mathbf{5.80}$ & 6.19 & $\mathbf{5.80}$ \\
          Fig. \ref{fig:local1}, Row 8 & 29.58 & 9.74 & 22.69 & $\mathbf{7.92}$ & 6.63 & $\mathbf{5.36}$ & 6.63 & $\mathbf{5.36}$ \\
          %\cdashlinelr{1-6}
      \bottomrule
  \end{tabular}
  \caption{A comparison of the optimization approach (with random initialization) and our hybrid method (with ``best of 9'' NN initialization) on a variety of challenging global and local image editing operations in Fig. \ref{fig:global1} and \ref{fig:local1}. The numbers indicate the RMSE of the outputs, and for reference, the first row showcases an input image that is reproducible by the shader.}
\label{tab:constrained-opt}
\end{table*}

\begin{table*}[ht]
\centering
\begin{tabular}{ccccccccccccccccc}
    \toprule
    & & \multicolumn{12}{c}{Image ID in sequence (i.e., $k$ of $\tilde{\myvec{t}}_{(k)}$)} \\
    \cmidrule(lr){3-15}
    F. evals & Technique & 0 & 10 & 20 & 30 & 40 & 50 & 60 & 70 & 80 & 90 & 100 & 110 & 120 & $\Sigma$\\
    \midrule
    \multirow{2}{*}{100} & No reinitialization & $\mathbf{1.93}$ & 1.67 & 2.19 & 2.90 & 3.82 & 4.79 & 5.73 & 6.81 & 7.93 & 9.14 & 10.43 & $\mathbf{11.55}$ & $\mathbf{12.99}$ & 81.88 \\
    & Reinitialization & $\mathbf{1.93}$ & $\mathbf{1.34}$ & $\mathbf{1.88}$ & $\mathbf{2.54}$ & $\mathbf{3.34}$ & $\mathbf{4.30}$ & $\mathbf{5.30}$ & $\mathbf{6.38}$ & $\mathbf{7.50}$ & $\mathbf{8.69}$ & $\mathbf{9.93}$ & $\mathbf{11.55}$ & $\mathbf{12.99}$ & $\mathbf{77.67}$ \\
    \cdashlinelr{1-16}
    \multirow{2}{*}{300} & No reinitialization & $\mathbf{1.64}$ & 1.47 & 2.07 & 2.80 & 3.70 & 4.62 & 5.70 & 6.75 & 7.86 & 9.00 & 10.21 & $\mathbf{11.41}$ & $\mathbf{12.82}$ & 80.05 \\
    & Reinitialization & $\mathbf{1.64}$ & $\mathbf{1.30}$ & $\mathbf{1.80}$ & $\mathbf{2.42}$ & $\mathbf{3.25}$ & $\mathbf{4.25}$ & $\mathbf{5.25}$ & $\mathbf{6.33}$ & $\mathbf{7.45}$ & $\mathbf{8.64}$ & $\mathbf{9.88}$ & $\mathbf{11.41}$ & $\mathbf{12.82}$ & $\mathbf{76.44}$ \\
    \cdashlinelr{1-16}
    \multirow{2}{*}{600} & No reinitialization & $\mathbf{1.57}$ & 1.44 & 2.06 & 2.77 & 3.66 & 4.60 & 5.69 & 6.74 & 7.83 & 8.96 & 10.12 & $\mathbf{11.41}$ & $\mathbf{12.80}$ & 79.65 \\
    & Reinitialization & $\mathbf{1.57}$ & $\mathbf{1.29}$ & $\mathbf{1.80}$ & $\mathbf{2.49}$ & $\mathbf{3.33}$ & $\mathbf{4.20}$ & $\mathbf{5.18}$ & $\mathbf{6.27}$ & $\mathbf{7.38}$ & $\mathbf{8.58}$ & $\mathbf{9.81}$ & $\mathbf{11.41}$ & $\mathbf{12.80}$ & $\mathbf{76.11}$ \\
    \bottomrule
 \end{tabular}
 \caption{Our proposed reinitialization technique consistently outperforms per-frame computation for the image sequence shown in Fig. \ref{fig:sequence}. The numbers indicate the RMSE of the outputs.}
\label{tab:animrender}
\end{table*}

\section{Results}
\label{sec:results}

In this section, we discuss the properties of our inverse problem formulation (i.e., inferring a shader setup that produces a prescribed input image), followed by both a quantitative and qualitative evaluation of our proposed hybrid method against the optimization and inversion network solutions. We also show that our system supports a wide variety of image editing operations and can rapidly predict image sequences. To ensure clarity, we briefly revisit the three introduced methods:
\begin{itemize}
\item The \textbf{optimization} approach relies on minimizing (\ref{eq:our_optimization}) with Nelder and Mead's simplex method using a random initial guess, and implementing $\phi$ through a neural renderer,
\item the \textbf{inversion network} refers to the ``best of 9'' inversion solution, i.e., $\myvec{x} \approx \phi_{(i)}^{-1}(\tilde{\myvec{t}})$ as shown in (\ref{eq:bestof9}),
\item our \textbf{hybrid method} is obtained by combining the two above approaches as described in Algorithm \ref{alg:pme}.
%minimizing (\ref{eq:our_optimization}) with the initial guess of $\myvec{x}_{init} = \phi^{-1}(\tilde{\myvec{t}})$.
\end{itemize}

Furthermore, in Appendix \ref{sec:appA}, we report the structure of the neural networks used to implement each individual $\phi_{(i)}^{-1}$ shown in Fig. \ref{fig:nn_fail}, and compare our solution to a selection of local and global minimizers in Appendix \ref{sec:appB}. At the end of this section, we also compare the total time taken to synthesize 1, 10, and 100 selected materials against a recent method for mass-scale material synthesis.
%(App. \ref{sec:appA}.). %and discuss how one should apportion the number of iterations and passes in (\ref{eq:our_optimization}) to obtain the best results (App. \ref{sec:appB}.). 

%\begin{figure}[htb!]
%\includegraphics[width=0.49\textwidth]{figures/plot2.png}
%\caption{Caption}
%\label{fig:plot2}
%\end{figure}

%\begin{figure*}[htb!]
%\includegraphics[width=1.0\textwidth]{figures/plots.png}
%\caption{Our hybrid solution outperforms the neural optimizer on capturing a simple reference material (left) and one of the more difficult local image stitching operations (right) and provides high-quality results within 20 seconds for each case.}
%\label{fig:plots}
%\end{figure*}

\textbf{Inversion accuracy.}
%As the forward operator $\phi$ is performed through neural rendering, its inverse can be implemented by training a neural network on the adjoint problem by swapping the inputs and outputs of the same dataset and adjusting the architecture appropriately. 
Our inversion technique leads to an approximate solution within a few milliseconds, however, because the structure of the forward and inverse networks differ, the inversion operation remains imperfect, especially when presented with a target image that includes materials that are only approximately achievable. To demonstrate this effect, we have trained 9 different inversion networks to implement $\phi^{-1}$ and show that none of the proposed solutions are satisfactory as a final output for the global colorization case (Fig. \ref{fig:nn_fail}). Our goal with this experiment was to demonstrate that a solution containing only one inversion network generally produces unsatisfactory outputs, regardless of network structure. However, these predictions can be used to equip our optimizer with an initial guess, substantially improving its results. As each neural network consumes between 300MB and 1GB of video memory, we were able to keep all of them loaded during the entirety of the work session. %Additionally, as each of these networks took roughly one second to load, it is possible to use our technique even in environments where video memory and GPU compute power is scarce.

%\begin{figure}[htb!]
%\includegraphics[width=0.49\textwidth]{figures/plot3.png}
%\caption{Caption}
%\label{fig:plot3}
%\end{figure}

\textbf{Optimizer and hybrid solution accuracy.}
In Table \ref{tab:constrained-opt}, we compared our hybrid solution against the ``best of 9'' inversion network and optimization approaches and recorded the RMS error after 50, 300 and 1500 function evaluations (these roughly translate to 1, 6, and 30-second execution times) to showcase the early and late-stage performance of these methods. The table contains a selection of scenarios that we consider to be the most challenging and note that the outputs showed no meaningful change after 1500 function evaluations. Our hybrid method produced the lowest errors in each of our test cases, and surprisingly, the inversion network initialization not only provides a ``headstart'' for our method, but also improves the final quality of the output, thereby helping the optimizer to avoid local minima. To validate the viability of our solutions, we also ran a global minimizer \cite{wales1997global} with several different parameter choices and a generous allowance of 30 minutes of computation time for each; our hybrid method was often able to match (and in some cases, surpass) the quality offered by this solution (Appendix \ref{sec:appB}, Table \ref{tab:opt-test}), further reinforcing how our inversion network initialization step helps avoid getting stuck in poor local minima. Note that the optimizer was unable to meaningfully improve the best prediction of the 9 inversion networks in Fig. \ref{fig:global1}, Row 7 -- in this case, a better solution can be found by using the prediction of only the first neural network and passing it to the optimizer, improving the reported RMSE from 22.38 to 19.39 by using 300 function evaluations.

\begin{figure}[tb!]
\includegraphics[width=0.4\textwidth]{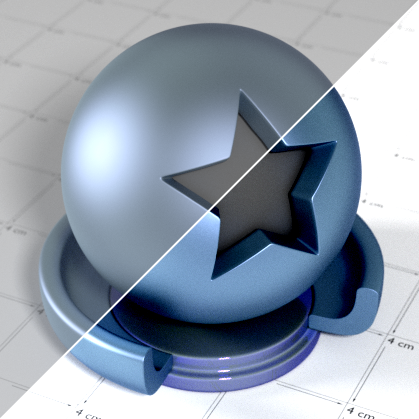}
\caption{Our image sequence starts with an input that is achievable using our shader (upper left), where each animation frame slightly increases its black levels. The lower right region showcases the 300th frame of the animation.}
\label{fig:sequence}
\end{figure}

\begin{figure}[htb!]
\includegraphics[width=0.49\textwidth]{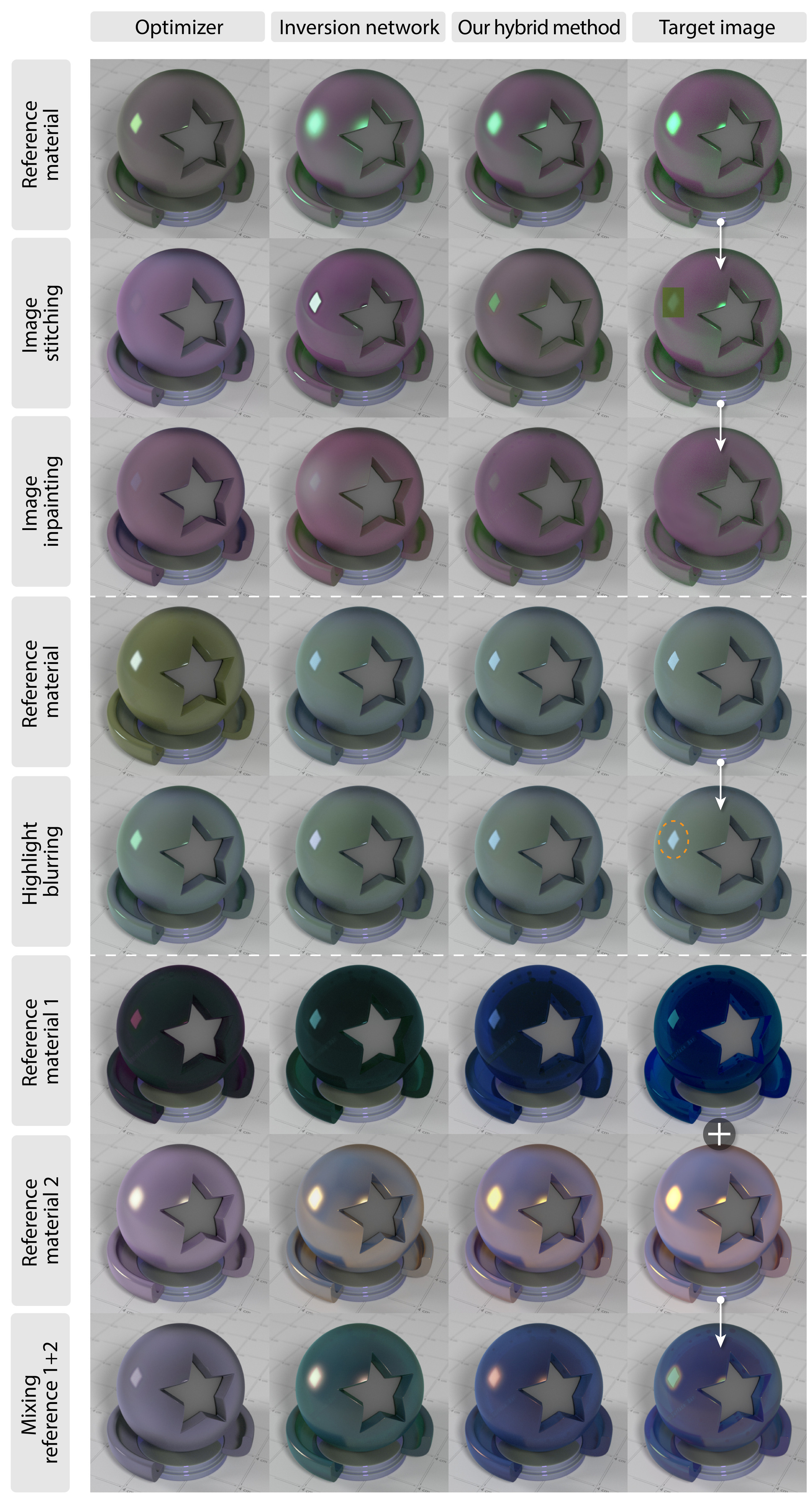}
\caption{Results for three techniques on local image editing operations and image mixing. The ``reference material'' labels showcase materials that can be obtained using our shader and are used as source images for the materials below them, where the arrows denote the evolution of the target image.}
\label{fig:local1}
\end{figure}

\textbf{Supported image editing operations.}
A typical workflow using our technique includes the artist choosing a source material and applying an appropriate image editing operation ($\Psi$) instead of engaging in a direct interaction with the principled shader. We cluster the set of possible transforms into \emph{global} (Fig. \ref{fig:global1}) and \emph{local} (Fig. \ref{fig:local1}) operations: these cases include saturation increase, grayscale transform, colorization, image mixing, stitching and inpainting, and selective blurring of highlights. Both the optimizer and our hybrid method were run for 1500 function evaluations to obtain the results showcased in these two figures. As these transformations come from a 2D raster editor and are not grounded in a physically-based framework, a perfect match is often not possible, however, in each of these cases, our hybrid method proposed a solution of equivalent or better quality compared to the ``best of 9'' inversion network and the optimizer solutions.

\textbf{Image sequence prediction.}
As our earlier results in Table \ref{tab:constrained-opt} revealed that the global colorization techniques typically prove to be among the more difficult cases, we have created a challenging image sequence with an input image that is achievable with our shader, and subjected it to a slight black level increase over many frames (Fig. \ref{fig:sequence}). Every image within this sequence is reproduced both with independent per-frame inference and our reinitialization technique with a strict time budget of 2, 6, and 12 seconds per image (100, 300, and 600 function evaluations). In Table \ref{tab:animrender}, we show that this simple extension successfully exploits the advantageous mathematical properties of these workflows and consistently reduces the output error for the majority of the sequence, i.e., images 1-100. We also report the RMSE of images 101-120 for reference, which we refer to as the ``converged'' regime in which the target images stray further and further away from the feasible domain, and the proposed solution remains the same despite these changes. Even in these cases, our reinitialization technique performs no worse than the ``no reinitialization'' method, and because of its negligible additional cost, we consider it to be a strictly better solution.

%Even as the utility of neural network predictions increases later in the image sequence (images 101-120), our reinitialization technique still performs no worse than the ``no reinitialization'' method, and due to its negligible additional cost, we therefore consider it to be a strictly better solution.

\begin{figure}[ht]
\includegraphics[width=0.48\textwidth]{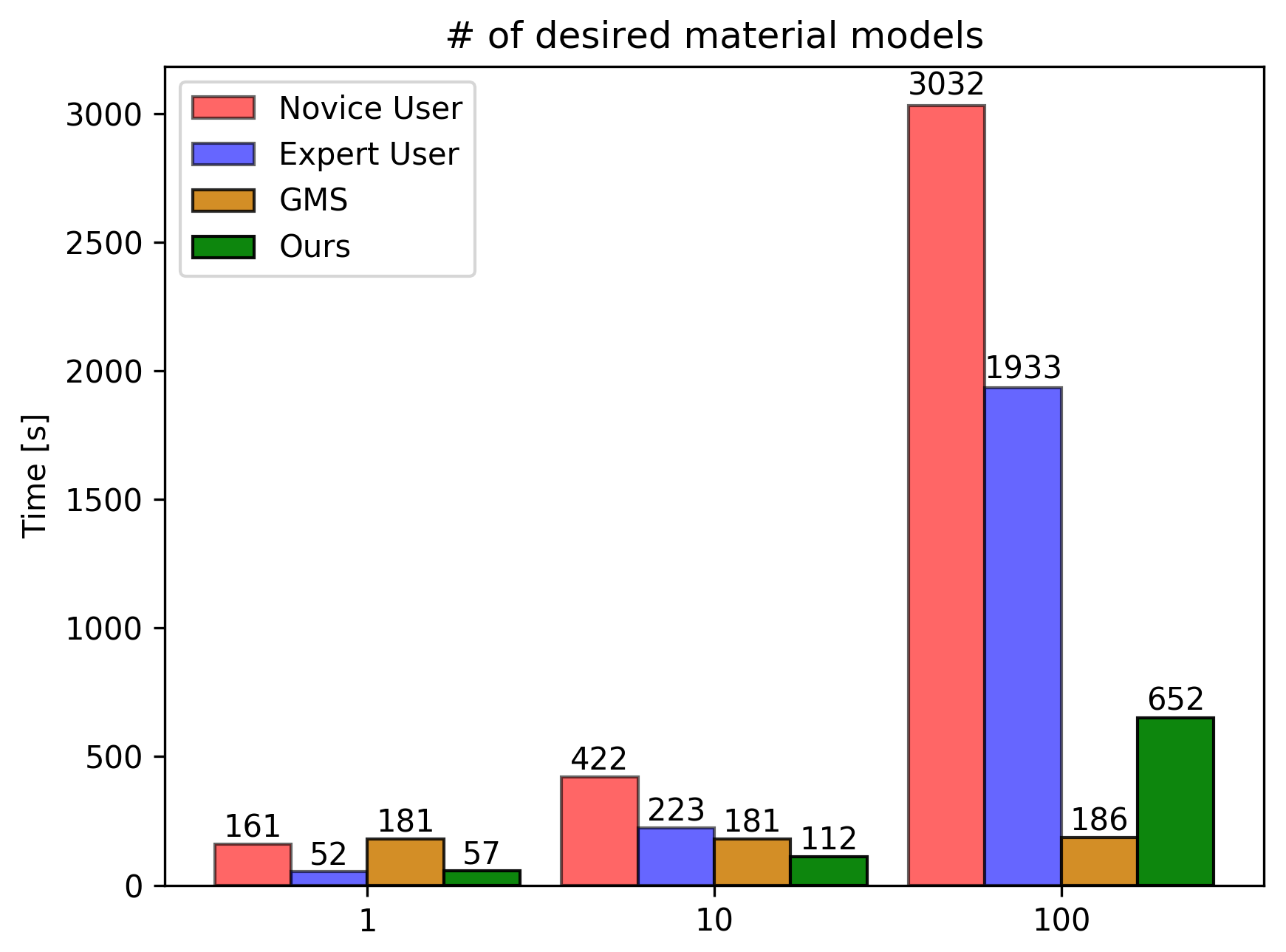}
\caption{The recorded modeling times reveal that if at most a handful (i.e., 1-10) of target materials are sought, our technique offers a favorable entry point for novice users into the world of photorealistic material synthesis.}
\label{fig:gms_comp}
\end{figure}

\textbf{Modeling and execution time.}
%The goal of our technique is to empower users without BSDF editing experience to create high-quality photorealistic materials. 
In Fig. \ref{fig:gms_comp}, we have recorded the modeling times for 1, 10, and a 100 similar materials using our method and compared them against Gaussian Material Synthesis \cite{Zsolnai18} (GMS), a learning-based technique for mass-scale material synthesis. We briefly describe the most important parameters of the recorded execution times and refer the interested reader to this paper for more details -- the novice and expert user timings were taken from the GMS paper and indicate the amount of time these users took to create the prescribed number of materials by hand using Disney's ``principled'' shader \cite{burley2012physically}, whereas GMS and our timings contain both the modeling (i.e., scoring a material gallery in GMS and performing image processing for our technique) and execution times. If only one material is desired, our technique outperforms this previous work and nearly matches the efficiency of an expert user. When 10 materials are sought (1 base material and 9 variants), our proposed method was adapted to use the re-initialization technique and offers the best modeling times, outperforming both GMS and expert users. In the case of mass-scale material synthesis, i.e., 100 or more materials, both methods outperform experts, where GMS offers the best scaling solution. In each case, the timings for our technique include the fixed cost of loading the 9 neural networks (5.5s). Throughout this manuscript, all results were generated using a NVIDIA TITAN RTX GPU.

% \begin{figure}[t!]
%     \centering
%     \begin{subfigure}[t]{0.16\textwidth}
%         \centering
%         \includegraphics{figures/sequence_target_images/0000-0-target.png}
%         \caption{}
%     \end{subfigure}%
%     ~ 
%     \begin{subfigure}[t]{0.16\textwidth}
%         \centering
%         \includegraphics{figures/sequence_target_images/0120-0-target.png}
%         \caption{}
%     \end{subfigure}
%     ~
%     \begin{subfigure}[t]{0.16\textwidth}
%         \centering
%         \includegraphics{figures/sequence_target_images/0300-0-target.png}
%         \caption{}
%     \end{subfigure}
%     \caption{Caption}
% \end{figure}

\section{Limitations and Future Work}
\label{sec:future}
As demonstrated in Fig. \ref{fig:nn_fail}, the results of $\phi^{-1}$ depend greatly on the performance of the encoder and decoder neural networks. As these methods enjoy significant research attention, we encourage further experiments in including these advances to improve them (e.g., architecture search \cite{real2017large}, capsule networks \cite{sabour2017dynamic,hinton2018matrix} and skip connections \cite{mao2016image} among many other notable works) and adapting other neural network architectures to our problem that are more tailored to solve inverse problems \cite{ardizzone2018analyzing,mataev2019deepred}. 
Furthermore, strongly localized edits, e.g., blurring a small part of a specular highlight typically introduces drastic changes within only a small subset of the image and represent only a small fraction of the RMSE calculations and thus may not get proper prioritization from the optimizer. To alleviate this, the relative importance of different regions may also be controlled via weighted masks to emphasize these edits, making these edited regions ``score higher'' in the error metric, offering the user more granular artistic control. In specialized cases, our reinitialization technique may prove to be useful for single images by using the parameter set used to produce $\myvec{t}$ as an initial guess for $\tilde{\myvec{t}}$. \Revision{In-scene editing still remains the key advantage of BRDF relighting techniques.}
%This is typically useful under the same conditions, i.e., when the two images are relatively close. 

We also note that our learning technique assumes an input shader of dimensionality $m$ and a renderer that is able to produce images of the materials that it encodes. In this work, our principled shader was meant to demonstrate the utility of this approach by showcasing intuitive workflows with the most commonly used BSDFs. However, this method needs not to be restricted to a classic principled BSDF, and is also expected to perform well on a rich selection of more specialized material models including thin-film interference \cite{dias1991ray,ikeda2015spectral}, fluorescence \cite{wilkie2001combined} birefringence \cite{weidlich2008realistic}, microfacet models \cite{heitz2016multiple} layered materials \cite{belcour:hal-01785457,Zeltner2018Layer}, and more.

\section{Conclusions}
\label{sec:conclusions}

\begin{figure*}[ht!]
\includegraphics[width=0.98\textwidth]{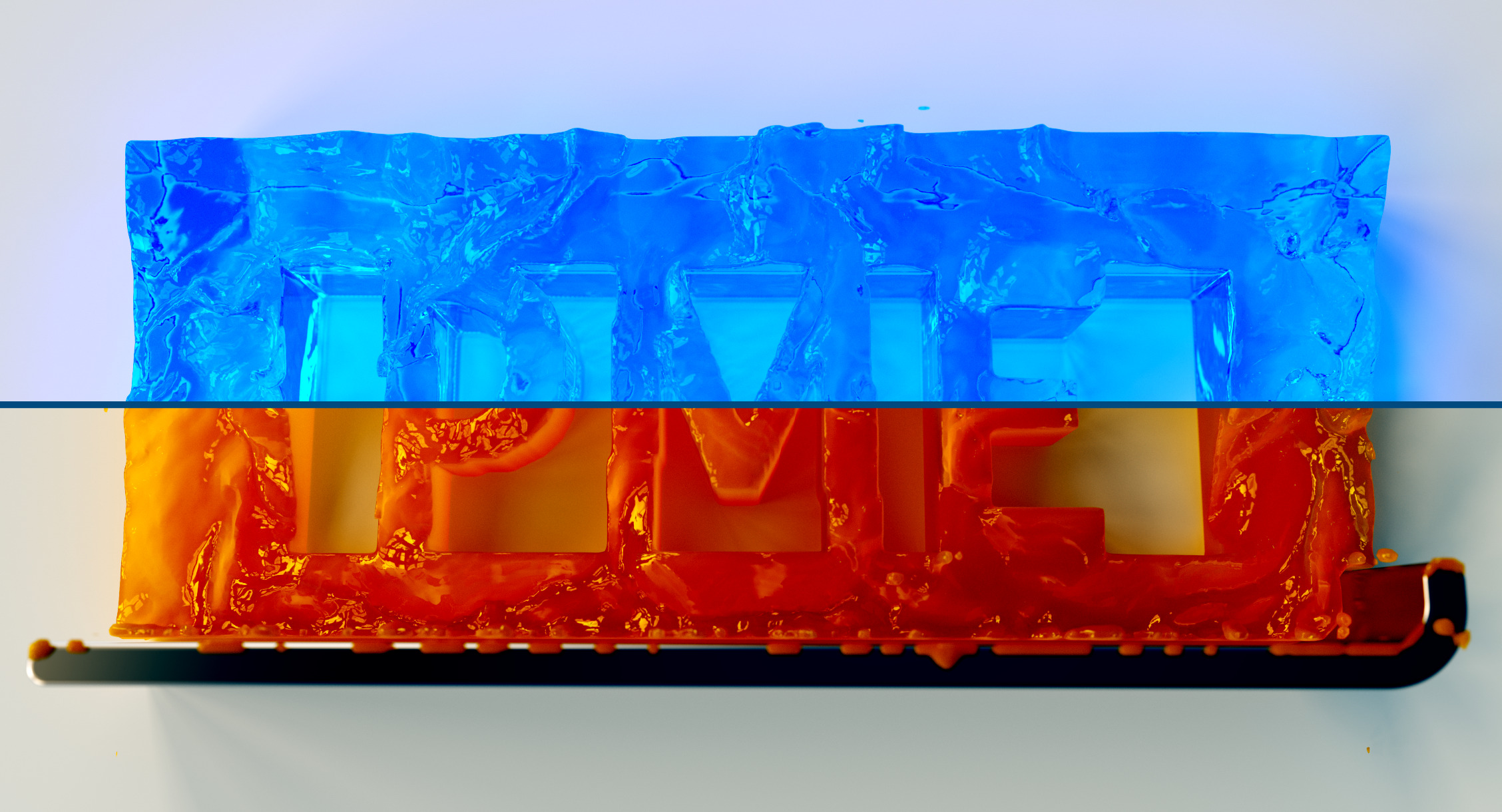}
\caption{Our technique is especially helpful early in the material design process where the user seeks to rapidly iterate over a variety of possible artistic effects. Both material types were synthesized using our described method. We demonstrate this workflow in our supplementary video.}
\label{fig:liquify_split}
\end{figure*}

We have presented a hybrid technique to empower novice users and artists without expertise in photorealistic rendering to create sophisticated material models by applying image editing operations to a source image. The resulting images are typically not achievable through photorealistic rendering, however, in many cases, solutions be found that are close to the desired output. Our learning-based technique is able to take such an edited image and propose a photorealistic material setup that produces a similar output, and provides high-quality results even in the presence of poorly-edited images. Our proposed method produces a reasonable initial guess and uses a neural network-augmented optimizer to fine-tune the parameters until the target image is matched as closely as possible. This hybrid method is simple, robust, and its computation time is within 30 seconds for every test case showcased throughout this paper. This low computation time is beneficial especially in the early phases of the material design process where a rapid iteration over a variety of competing ideas is an important requirement (Fig \ref{fig:liquify_split}). 
\Revision{Our two key insights can be summarized as follows:
\begin{itemize}
	\item Normally, using an input image that was generated by a principled shader is not useful given that the user has to generate this image themselves with a known parameter setup. However, our main idea is that the user can subject this image to raster editing operations and ``pretend'' that this input is achievable, and reliably infer a shader setup to mimic it.
	\item Our neural networks can be combined with optimizers both \emph{directly}, i.e., by using an optimizer that invokes a neural renderer at every function evaluation step to speed up the convergence and \emph{indirectly} by using a set of neural networks network to endow the optimizer with a reasonable initial guess (steps \ding[1.15]{184} and \ding[1.15]{185} in Fig. \ref{fig:workflow}). This combination results in a two-stage sytem that opens up efficient material editing workflows for artists without expertise in this area.
\end{itemize}}
Furthermore, we proposed a simple extension to support predicting image sequences with a strict time budget of 1-2 seconds and believe this method will offer an appealing entry point for novices into world of photorealistic material modeling.

\Revision{
\section*{Acknowledgments}
We would like to thank Reynante Martinez for providing us the geometry and some of the materials for the Paradigm (Fig. \ref{fig:teaser}) and Genesis scenes (Fig. \ref{fig:glass-scene}), ianofshields for the Liquify scene that served as a basis for Fig. \ref{fig:liquify_split}, Robin Marin for the material test scene, Andrew Price and G\'{a}bor M\'{e}sz\'{a}ros for their help with geometry modeling, Fel\'{i}cia Zsolnai-Feh\'{e}r for her help improving our figures, Christian Freude, David Ha, Philipp Erler and Adam Celarek for their useful comments. We also thank NVIDIA for providing the hardware to train our neural networks. This work was partially funded by Austrian Science Fund (FWF), project number P27974.}
\appendix
\section{Neural network architectures}
\label{sec:appA}

Below, we describe the neural network architectures we used to implement $\phi_{(i)}^{-1}$. The Conv2D notation represents a 2D convolutional layer with the appropriate \emph{number of filters}, \emph{spatial kernel sizes} and \emph{strides}, where FC represents a dense, fully-connected layer with a prescribed number of \emph{neurons} and \emph{dropout probability}.

\begin{enumerate}
	%$4$x$\{$Conv1D($64,3,1$) -- Upsampling($2$)$\}$ -- FC($1000$) -- FC($),
	\item 2x$\{$Conv2D(32,3,1), MaxPool(2,2)$\}$ -- \newline 1x$\{$Conv2D(64,3,1), MaxPool(2,2)$\}$ -- \newline 2x$\{$Conv2D(128,3,1), MaxPool(2,2)$\}$ -- \newline 2x$\{$FC(1000, 0.1)$\}$ - FC(m, 0.0)
	\item 2x$\{$Conv2D(32,3,1), MaxPool(2,2)$\}$ -- \newline 2x$\{$FC(1000, 0.1)$\}$ - FC(m, 0.0)
	\item 2x$\{$Conv2D(32,3,1), MaxPool(2,2)$\}$ -- \newline 2x$\{$FC(1000, 0.5)$\}$ - FC(m, 0.0)
	\item 2x$\{$Conv2D(32,3,1), MaxPool(2,2)$\}$ -- \newline 1x$\{$Conv2D(64,3,1), MaxPool(2,2)$\}$ -- \newline 2x$\{$Conv2D(128,3,1), MaxPool(2,2)$\}$ -- \newline 2x$\{$FC(3000, 0.5)$\}$ - FC(m, 0.0)
	\item 2x$\{$Conv2D(32,3,1), MaxPool(2,2)$\}$ -- \newline 1x$\{$Conv2D(64,3,1), MaxPool(2,2)$\}$ -- \newline 2x$\{$Conv2D(128,3,1), MaxPool(2,2)$\}$ -- \newline 2x$\{$FC(3000, 0.0)$\}$ - FC(m, 0.0)
	\item 2x$\{$Conv2D(32,3,1), MaxPool(2,2)$\}$ -- \newline 2x$\{$FC(1000, 0.0)$\}$ - FC(m, 0.0)
	\item 2x$\{$Conv2D(32,3,1), MaxPool(2,2)$\}$ -- \newline 2x$\{$FC(1000, 0.0)$\}$ - FC(m, 0.0)
	\item 2x$\{$Conv2D(32,3,1), MaxPool(2,2)$\}$ -- \newline 2x$\{$FC(100, 0.0)$\}$ - FC(m, 0.0)
	\item 2x$\{$Conv2D(32,3,1), MaxPool(2,2)$\}$ -- \newline 2x$\{$FC(1000, 0.0)$\}$ - FC(m, 0.0)
\end{enumerate}

Neural networks 6,7 and 9 are isomorphic and were run for a different number of epochs to test the effect of overfitting later in the training process, and therefore offer differing validation losses. The implementation of $\phi$ is equivalent to the one used in Zsolnai-Feh\'{e}r et al.'s work \shortcite{Zsolnai18}.

\section{Comparison of optimizers}
\label{sec:appB}

In Table \ref{tab:opt-test}, we have benchmarked several optimizers, i.e., L-BFGS-B \cite{byrd1995limited}, SLSQP \cite{kraft1994algorithm}, the Conjugate Gradient method \cite{hestenes1952methods} and found Nelder and Mead's simplex-based self-adapting optimizer \shortcite{nelder1965simplex} to be the overall best choice for our global and local image-editing operations. For reference, we also ran Basin-hopping \cite{wales1997global}, a global minimizer with a variety of parameter choices and a generous allowance of 30 minutes of execution time for each test case. This method is useful for challenging non-linear optimization problems with high-dimensional search spaces. Note that when being run for long enough, this technique is less sensitive to initialization due to the fact that it performs many quick runs from different starting points, and hence, we report one result for both initialization techniques. The cells in the intersection of ``Nelder-Mead'' and ``NN'' denote our proposed hybrid method, which was often able to match, and in some cases, outperform this global minimization technique.

\begin{table*}[htb]
\centering
	\begin{tabular}{cccccccc}
	  \toprule
	  Input & Init. type & Init. RMSE & Nelder-Mead & L-BFGS-B & SLSQP & CG & Basin-hopping \\
	  \midrule
	  Fig. \ref{fig:global1}, Row 1 & Rand & 41.93 & 5.62 & 20.47 & 17.96 & 5.24 & \multirow{2}{*}{\textbf{2.01}} \\
	  Fig. \ref{fig:global1}, Row 1 & NN & 5.94 & 2.37 & 5.84 & \color{red}{5.94} & \color{red}{5.94} &  \\
	  \cdashlinelr{1-8}
	  Fig. \ref{fig:global1}, Row 2 & Rand & 78.45 & 40.21 & \color{red}{78.45} & \color{red}{78.45} & \color{red}{78.45} & \multirow{2}{*}{\textbf{32.67}} \\
      Fig. \ref{fig:global1}, Row 2 & NN & 32.72 & \textbf{32.67} & \color{red}{32.72} & \color{red}{32.72} & \color{red}{32.72} & \\
      \cdashlinelr{1-8}
      Fig. \ref{fig:global1}, Row 4 & Rand & 35.37 & 16.98 & 28.84 & \color{red}{35.37} & 34.99 & \multirow{2}{*}{14.72} \\
      Fig. \ref{fig:global1}, Row 4 & NN & 18.68 & \textbf{14.68} & 15.33 & 18.18 & 15.90 & \\
      \cdashlinelr{1-8}
      Fig. \ref{fig:global1}, Row 7 & Rand & 41.65 & 26.24 & \color{red}{41.65} & \color{red}{41.65} & \color{red}{41.65} & \multirow{2}{*}{\textbf{22.38}} \\
      Fig. \ref{fig:global1}, Row 7 & NN & 22.42 & \textbf{22.38} & \color{red}{22.42} & \color{red}{22.42} & \color{red}{22.42} & \\
      \cdashlinelr{1-8}
      Fig. \ref{fig:global1}, Row 8 & Rand & 29.04 & 22.93 & \color{red}{29.04} & 26.71 & 28.21 & \multirow{2}{*}{15.69} \\
      Fig. \ref{fig:global1}, Row 8 & NN & 19.82 & \textbf{15.37} & \color{red}{19.82} & \color{red}{28.87} & \color{red}{19.82} & \\
      \cdashlinelr{1-8}
      Fig. \ref{fig:local1}, Row 2 & Rand & 23.78 & 8.26 & \color{red}{23.78} & \color{red}{23.78} & 21.75 & \multirow{2}{*}{\textbf{7.63}} \\
      Fig. \ref{fig:local1}, Row 2 & NN & 12.79 & 7.80 & \color{red}{12.79} & \color{red}{12.79} & \color{red}{12.79} & \\
      \cdashlinelr{1-8}
      Fig. \ref{fig:local1}, Row 3 & Rand & 21.60 & 6.19 & \color{red}{21.60} & \color{red}{21.60} & 20.83 & \multirow{2}{*}{5.86} \\
      Fig. \ref{fig:local1}, Row 3 & NN & 9.09 & \textbf{5.80} & \color{red}{9.09} & \color{red}{9.09} & \color{red}{9.09} & \\
      \cdashlinelr{1-8}
      Fig. \ref{fig:local1}, Row 8 & Rand & 29.58 & 6.63 & \color{red}{29.58} & \color{red}{29.58} & \color{red}{29.58} & \multirow{2}{*}{\textbf{5.07}} \\
      Fig. \ref{fig:local1}, Row 8 & NN & 9.74 & 5.36 & 9.61 & 9.61 & 9.68 & \\
      %\cdashlinelr{1-8}
		  \bottomrule
	\end{tabular}
	\caption{A comparison of a set of classical optimization techniques revealed that when using Nelder and Mead's simplex-based optimizer with our ``best of 9'' inversion network initialization, we can often match, and in some cases, outperform the results of Basin-hopping, a global minimizer. In the interest of readability, we have marked the cases where the optimizers were unable to improve upon the initial guess with red. For reference, the first two rows showcase an input image that is reproducible by the shader.}
\label{tab:opt-test}
\end{table*}

%\section{Efficient use of iterations}
%\label{sec:appB}

\bibliographystyle{ACM-Reference-Format}
\bibliography{sample-bibliography} 

\end{document}